\documentclass[%
reprint,
 amsmath,amssymb,
 aps,superscriptaddress
]{revtex4-2}

\usepackage{graphicx}% Include figure files
\usepackage{dcolumn}% Align table columns on decimal point
\usepackage{bm}% bold math
\usepackage{hyperref}% add hypertext capabilities
\usepackage{color}
\newcommand{\comment}[1]{}

\begin{document}

%\preprint{APS/123-QED}

\title{Higgs and Goldstone spin-wave modes in striped magnetic texture}% Force line breaks with \\

\author{Mat\'{\i}as Grassi}
\affiliation{%
 Universit\'e de Strasbourg, CNRS, Institut de Physique et Chimie des Mat\'eriaux de Strasbourg, UMR 7504, F-67000 Strasbourg, France.
}%

\author{Moritz Geilen}
\affiliation{%
Fachbereich Physik and Landesforschungszentrum OPTIMAS, Technische Universit\"{a}t Kaiserslautern, 67663 Kaiserslautern, Germany.
}%

\author{Kosseila Ait Oukaci}
\affiliation{%
 Institut Jean Lamour, Universit\'e de Lorraine, UMR 7198, CNRS, F-54000 Nancy, France.
}%

\author{Yves Henry}
\affiliation{%
 Universit\'e de Strasbourg, CNRS, Institut de Physique et Chimie des Mat\'eriaux de Strasbourg, UMR 7504, F-67000 Strasbourg, France.
}%

\author{Daniel Lacour}
\affiliation{%
 Institut Jean Lamour, Universit\'e de Lorraine, UMR 7198, CNRS, F-54000 Nancy, France.
}%

\author{Daniel Stoeffler}
\affiliation{%
Universit\'e de Strasbourg, CNRS, Institut de Physique et Chimie des Mat\'eriaux de Strasbourg, UMR 7504, F-67000 Strasbourg, France.
}%

\author{Michel Hehn}
\affiliation{%
 Institut Jean Lamour, Universit\'e de Lorraine, UMR 7198, CNRS, F-54000 Nancy, France.
}%

\author{Philipp Pirro}
\affiliation{%
 Fachbereich Physik and Landesforschungszentrum OPTIMAS, Technische Universit\"{a}t Kaiserslautern, 67663 Kaiserslautern, Germany.
}%

\author{Matthieu Bailleul}
\affiliation{%
 Universit\'e de Strasbourg, CNRS, Institut de Physique et Chimie des Mat\'eriaux de Strasbourg, UMR 7504, F-67000 Strasbourg, France.
}%

\date{\today}% It is always \today, today,

\begin{abstract}
Spontaneous symmetry breaking is ubiquitous in physics. Its spectroscopic signature consists in the softening of a specific mode upon approaching the transition from the high symmetry side and its subsequent splitting into a zero-frequency ‘Goldstone’ mode and a non-zero-frequency ‘Higgs’ mode. Although they determine the whole system dynamics, these features are difficult to address in practice because of their vanishing coupling to most experimental probes and/or their strong interaction with other fluctuations. In this work, we consider a periodic magnetic modulation occurring in a ferromagnetic film with perpendicular-to-plane magnetic anisotropy and observe its Goldstone and Higgs spin-wave modes at room temperature using microwave and optical techniques. This simple system constitutes a particularly convenient platform for further exploring the dynamics of symmetry breaking.
%\begin{description}
%\end{description}
\end{abstract}

\maketitle

%\tableofcontents

\twocolumngrid
\section{\label{sec:level1}Introduction}

Upon spontaneous symmetry breaking, a system organizes itself in a state with a lower symmetry than that of its constituting entities, as exemplified by superconducting, magnetic or incommensurate structural phases~\cite{1,2}. According to the Landau theory, such transition is conveniently visualized by defining an order parameter $\psi$ and following the morphology of the free energy surface $E(\psi)$~\cite{3}. This is illustrated in Fig. 1(a) for a system with $U(1)$ symmetry, where $\psi$ is a complex number (or, equivalently, a two-dimensional real vector). In the high symmetry phase, $E$ presents a single minimum at $\psi=0$. Upon driving the system through the transition, the curvature around this point decreases, reaches zero at the critical point and then changes sign. In the low symmetry phase, the energy surface eventually takes the shape of a Mexican hat with a degenerate minimum extending over a circle of radius $|\psi|=\psi_0$. The system has to ``choose'' a phase arg$(\psi)$, which constitutes the symmetry break. This particular energy landscape gives rise to characteristic low frequency dynamic modes, conveniently viewed as the oscillations of a mass moving on such surface~\cite{1}. Upon driving the system from the high symmetry phase, the oscillations around the $\psi=0$ minimum [blue arrow in Fig. 1(a)] are expected to soften gradually, reach zero frequency at the critical point, and subsequently splits in two, a zero-frequency mode with azimuthal trajectory along the rim and a non-zero frequency mode with radial trajectory across the rim (solid and dashed red arrows in Fig. 1(a), respectively). These two modes, referred as Goldstone and Higgs modes, respectively, dominate the whole dynamics of the low symmetry phase, but also its coupling to external degrees of freedom, in particular gauge ones. Originally explored in the context of superconductivity, the latter is of particular importance for particle-physics, as the finite masses of the W and Z weak bosons can only be explained by their coupling to the symmetry-breaking Higgs field~\cite{4,5}. 

Although these dynamics are of crucial interest, their direct observation is a serious challenge as it requires to drive the system through the transition while keeping experimental access to the relevant low frequency excitations, particularly prone to overdamping due to defects and thermal/quantum microscopic fluctuations~\cite{2,6}. This difficulty could be avoided in low-temperature inelastic neutron scattering studies of very specific pressure-induced structural and magnetic transitions~\cite{7,8}. More recently, it was proposed to use artificial systems, namely ultra-cold bosons lattices, which can be driven through a quantum phase transition, their excitations being characterized via real time optical spectroscopy~\cite{9,10}.  

In this article, we show that an archetypal system of micromagnetism, the so-called magnetic weak stripes, allows for a room-temperature observation of the Higgs/Goldstone dynamics by inelastic light scattering and microwave spectroscopy. Magnetic weak stripes consist of a field-tunable modulation at a mesoscopic scale occurring in ferromagnetic films possessing a moderate perpendicular magnetic anisotropy~\cite{12}. The stability of this texture was predicted in the early days of theoretical micromagnetism \cite{3, Landau1935,Muller1961, Brown1961}, and later confirmed by static magnetic imaging~\cite{Spain1963, Saito1964}. Several studies have explored the corresponding dynamics evidencing a complex set of vibration modes varying to a very large extent with the magnetic parameters of the film and upon application of a control magnetic field. Initially described within local resonance models (domain/domain-wall-resonance)~\cite{22,23,24}, this complex phenomenology has recently been rephrased in the vocabulary of magnonics, as a complex set of spin-wave modes localized/scattered by the periodic modulation~\cite{18,25,28}. Moving ahead in that direction, we provide here a unified description of both statics and dynamics of magnetic stripes based on the identification of a specific spin-wave mode which, upon stripe nucleation, softens and then splits into a Goldstone/Higgs pair. For this purpose, we first formulate an elementary analytical model of the stripe statics and related spin-wave dynamics based on the Landau theory of phase transitions. Then, we report inelastic light scattering measurements conducted down to spin-wave wavelength of the order of the modulation period for different points across the critical region. Confronting them with complementary ferromagnetic resonance measurements and micromagnetic simulations, we arrive at an unprecedented global picture of the low frequency dynamics related to such symmetry breaking.

%In this report, we describe another path allowing for a room-temperature observation of the Higgs/Goldstone dynamics and a particularly explicit model of it. We use a ferromagnetic film hosting a field-tuneable modulation at a mesoscopic scale of a few hundreds of nm, which makes it relatively insensitive to microscopic fluctuations and allow for a convenient probing of its spin-wave dynamics by inelastic light scattering and microwave spectroscopy techniques. 

%Fig1
\begin{figure*}
	\centering
	\includegraphics[width=0.8\linewidth]{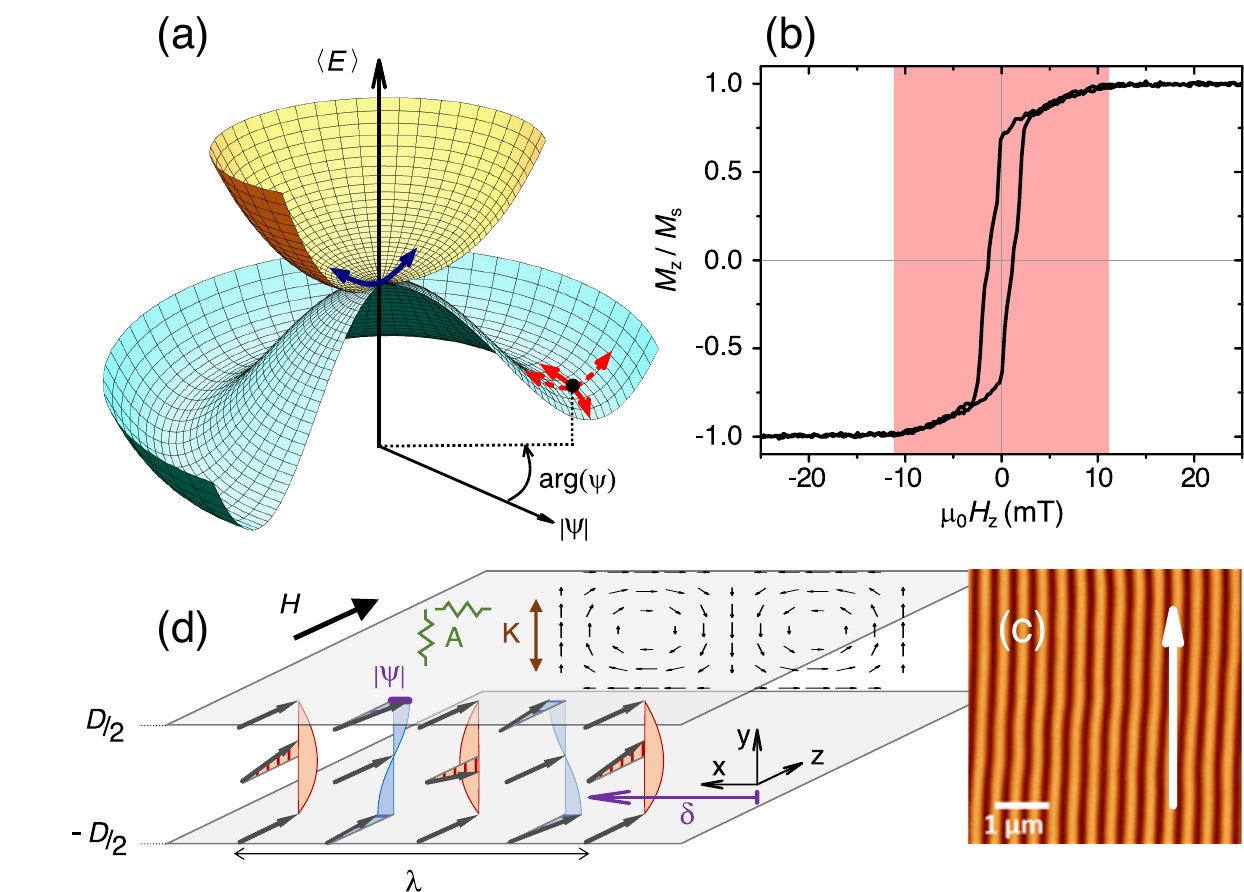}
	\caption{(a) Sketch of the characteristic dynamic modes associated with a $U(1)$ symmetry breaking. The blue arrow shows the degenerate modes in the high-symmetry phase (yellow potential surface). The solid and dashed red arrows show the Goldstone and Higgs modes in the low-symmetry phase (sombrero-shape blue potential surface). (b) Magnetization loop measured for our $180~$nm Co$_{40}$Fe$_{40}$B$_{20}$ film with a magnetic field $\bm{H}$ in the plane. (c) Magnetic force microscopy image of the weak stripes magnetic texture existing below the critical field $H_c$ (shown here at remanence). (d) Sketch of the three-dimensional normalized magnetization distribution  $\bm{M}(x,y,z)/M_S$ within the weak stripes texture, together with its minimal description in terms of a U(1) symmetry breaking, with associate amplitude $|\psi|$ and phase arg$(\psi)=k\,\delta$ (see details in the text). }\label{Fig_1}
\end{figure*}

\section{\label{sec:level1}Results}

The system studied consists of an amorphous Co$_{40}$Fe$_{40}$B$_{20}$ film of thickness $D = 180~$nm deposited on intrinsic silicon and initialized by applying a saturating magnetic field $\bm{H}$ in the film plane. Upon reducing the magnitude of the field below about $12~$mT, the average magnetization of the film starts to decrease in a roughly linear fashion [Fig. 1(b)]. In this regime, magnetic force microscopy allows one to identify a modulation periodic in one-dimension with a wavenumber of about $2\pi/(300$~nm$)=21~$rad/$\mu$m [Fig. 1(c)]~\cite{11}, identified with the archetypical ‘magnetic weak stripes’ arising in films which possess a moderate perpendicular magnetic anisotropy $K$~\cite{12}. In the following, we shall revisit this texture from the point of view of its dynamics. For this purpose, we start by describing a minimal model for stripe nucleation, as sketched in Fig. 1(d). The mechanism consists of a competition between the in-plane magnetic field $\bm{H}$, which tends to maximize the component of the magnetization distribution $\bm{M}(x,y)$ along its direction $\bm{\hat{z}}$ (the system is assumed to be invariant along $z$), and the perpendicular magnetic anisotropy, which tends to maximize its (out-of-plane) $y$ component. The inhomogeneity of the texture is induced by the dipolar interaction. In order to avoid the large demagnetizing energy density that would be associated with a uniform out-of-plane excursion of magnetization ($\frac{\mu_0}{2} M_y^2$, $\mu_0$ being the permeability of vacuum), the latter favors an alternation of sign of $M_y$ in the film interior, along the transverse direction $x$, together with a closure of the resulting magnetic flux via quadrature sign changes of the transverse component $M_x$ at both film surfaces. Finally, the overall distribution is smoothed out by the exchange energy density $A\nabla^2 \bm{M}$, where $A$ is the exchange stiffness constant. For the magnetic parameters of our film ($M_S = 1330~$kA/m, $K = 32.7~$kJ/m$^3$ $A = 16.6~$pJ/m)~\cite{Supp} the comparison of the different energy scales deduced from a dimensional analysis $\mu_0 M_S^2 >> K \sim \frac{A}{D^2}$ suggests the use of a stray-field-free ansatz of the magnetization distribution, which cancels the dominant demagnetizing energy while reducing the magnetic anisotropy contribution with respect to an in-plane saturated state. Following Hubert~\cite{13}, we write this ansatz as a combination of two sinusoidal functions in quadrature with each other:

\begin{eqnarray}
\label{Ansatz}
\frac{\bm{M}}{M_S}(x,y)=&&|\psi|{\Big\lbrace}  \text{sin}\left[k(x-\delta)\right]\text{sin}\left(\frac{\pi y}{D}\right) \bm{\hat{x}}+\nonumber\\
&&\frac{k D}{\pi} \text{cos}\left[k(x-\delta)\right]\text{cos}\left(\frac{\pi y}{D}\right)\bm{\hat{y}}{\Big\rbrace}.
\end{eqnarray}

This ensures the vanishing of both surface magnetic charges ($M_y (y=\pm D/2)=0$) and volume magnetic charges ($\frac{\partial M_x}{\partial x}+\frac{\partial M_y}{\partial y}=0$). Here $k$ is the wavenumber of the modulation ($k =2\pi/\lambda$ for which we will select later the value minimizing the total energy). With the two other parameters defining the modulation, namely its amplitude $|\psi|$ (taken as the maximum in-plane excursion of the magnetization) and its lateral positioning $\delta$ [measured with respect to an arbitrary reference point, see Fig. 1(d)], we build a complex number $\psi=|\psi|e^{ik\delta}$ that we identify with the order parameter of the stripe texture. Then, following Landau theory, ~\cite{3} we develop the spatially averaged magnetic energy density $\left\langle E\right\rangle=E_0+a(k,H)|\psi|^2+b(k,H)|\psi|^4+O(|\psi|^6)$, the terms with odd powers being zero by symmetry (see~\cite{Supp} for details). This simple form allows us to derive an analytical estimate of the field and wavenumber at the critical point ($\mu_0 H_c=\frac{2\,K}{M_S}-\frac{4\,\pi\sqrt{A\,K}}{D\, M_s} =10.5~$mT, $k_c=\frac{\pi}{D}\sqrt{\frac{2\,K+\mu_0 M_S H_c}{2\,K-\mu_0\, M_S\, H_c}}=21.6~$rad/$\mu$m, respectively, as deduced from the conditions $a=\frac{\partial a}{\partial k}=0$), and the modulation amplitude below nucleation $\psi_0=\sqrt{\frac{-a(k_c,H)}{2\,b(k_c,H)}}$ . Despite its simplicity, this explicit model captures most of the physics of the weak stripes observed experimentally. Despite a small underestimate of  critical field of $1.3~$mT, it is also in good agreement with micromagnetic simulations (see~\cite{Supp}).

%Fig2
\begin{figure}
	\centering
	\includegraphics[width=1\linewidth]{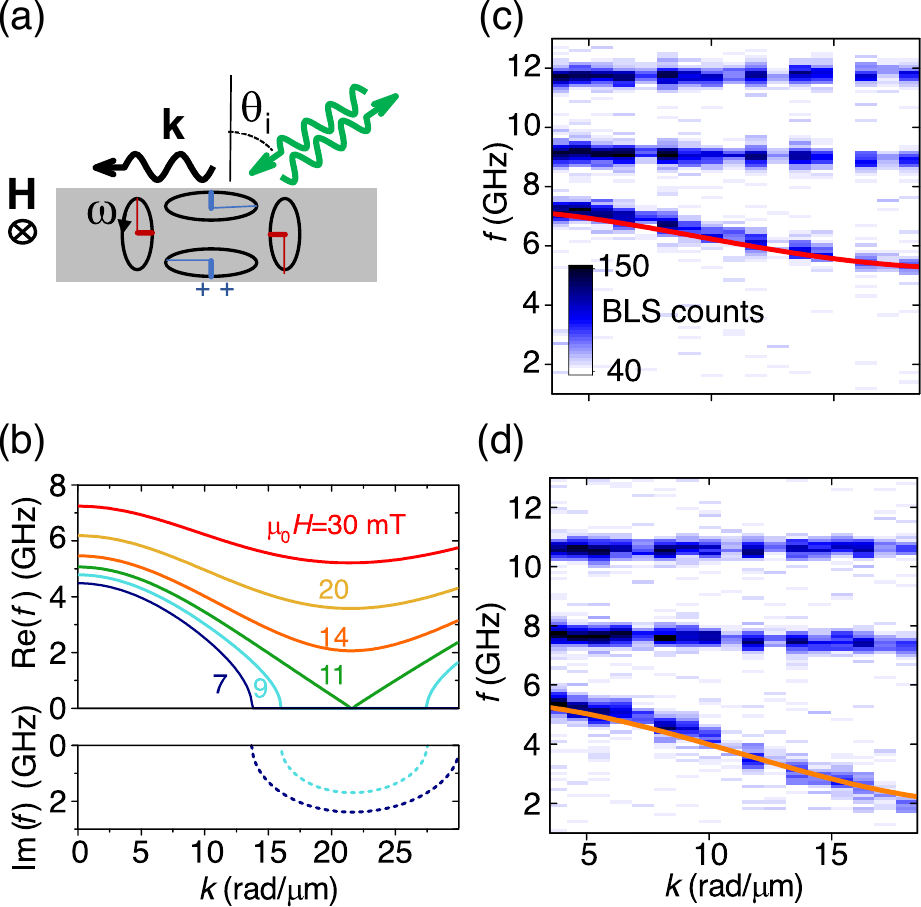}
	\caption{(a) Sketch of the Damon-Eshbach geometry, magnetization precession (black ellipses) and basis functions used for modelling (red and blue bars). The thin bars can be combined to form the stray-field-free ansatz of Eq. (1), while the thick bars, orthogonal to them, carry non-zero magnetic pseudo-charges, as shown at the bottom of the film. (b) Spin-wave dispersion relations calculated analytically using the basis of panel (a) (see details in the text and in~\cite{Supp}), for an external field of 30, 20, 14, 11, 9 and $7~$mT. For the last two values, the saturated state is unstable, and the spin-wave frequency becomes imaginary over a certain range of wavenumbers (see the distinct vertical scale at the bottom of the graph). (c,d), Color maps of the Brillouin light scattering intensity measured as function of the transferred wave-vector (scattering geometry shown by the green arrows in panel (a) and transferred frequency $f$  under an external field $\mu_0 H=30$ and $14~$mT, respectively. The lines show the calculated soft mode frequency [same as in panel (b)].}\label{Fig_2}
\end{figure}

Our minimal model of stripe nucleation forms the skeleton of a description of the spin wave dynamics in this regime: We place ourselves in the saturated state and consider a plane-wave of angular frequency $\omega$ and wavenumber $k$ propagating along $\bm{\hat{x}}$  , $\bm{m}(y)\,e^{i(\omega t-kx)}$  [spin-wave configuration referred to as Damon-Eshbach, Fig. 2(a)]~\cite{14}. Its complex amplitude distribution $\bm{m}(y)$ is written as a linear combination of four vector functions : the two functions sin$(\pi y/D)\bm{\hat{x}}$ and cos$(\pi y/D)\bm{\hat{y}}$  appearing in the static ansatz of Eq.(1) and two extra ones sin$(\pi y/D)\bm{\hat{y}}$ and cos$(\pi y/D)\bm{\hat{x}}$ obtained through a local 90$^\circ$ rotation and necessary for describing the precession of the magnetization. Unlike the former, the latter pair of functions carries magnetic pseudo-charges [Fig. 2(a)], so that the precession will lead to sizeable stray fields. We identify the associated demagnetizing energy with a kinetic energy, which, combined with the potential energy $\left\langle E\right\rangle$ described above, will determine the mode frequency, in analogy with the D{\"o}ring mass term of magnetic domain-wall dynamics~\cite{12}. More specifically, we shall project the linearized equation of motion of the magnetization, $i\omega \bm{m}=\gamma M_S\bm{\hat{z}}\times \frac{\partial E}{\partial \bm{m}}$, where $\gamma$ is the gyromagnetic ratio, onto this basis set and diagonalize the resulting 4$\times$4 matrix (Eq. S9) to obtain eigenfrequencies and eigenmodes~\cite{15}. The spin-wave dispersion relation $\omega(k)/(2\,\pi)$ obtained for the lowest frequency mode is shown in the top panel of Fig. 2(b) for different values of the applied field. Far above nucleation, one distinguishes clearly a non-monotonic wave-vector dependence with a minimum frequency at a non-zero wave vector of about $21~$rad/$\mu$m. This minimum constitutes the dynamic precursor of stripe nucleation: its wavenumber is the critical one $k_c$ of the stripe modulation and  its frequency tends to zero as $H$ approaches $H_c$. This allows us to reinterpret stripe domain nucleation as the freezing of the lowest frequency spin-wave of the system~\cite{16}. 

To observe this characteristic mode softening, we now resort to Brillouin light scattering (BLS), an inelastic light scattering technique capable of probing thermally excited spin waves over a broad range of wave-vectors. The measurement geometry is sketched in Fig. 2(a):  the film is illuminated with a laser beam under an angle of incidence $\theta_i$ in the presence of a magnetic field $\bm{H}$ perpendicular to the incidence plane. The backscattered light is collected and frequency-analyzed with a high finesse Fabry-P\'erot interferometer. Due to the conservation of energy and in-plane linear momentum, the frequency shift of the scattered light and the transferred wave-vector ($k =4\pi/\lambda_{\text{laser}}\,\text{sin}(\theta_i)$) are to be identified with those of the quasi-particles absorbed/emitted during the scattering process. In our case, these are the spin-waves which couple to light via magneto-optical effects~\cite{17}. 

Fig. 2(c) shows a color plot of the spectra recorded in the saturated state with $\mu_0 H = 30~$mT for different transferred wave vector, thus providing a direct picture of the spin-wave dispersions, up to a wave-vector of about $k = 18~$rad/$\mu$m. One recognizes clearly three spin-wave branches. The two highest ones with nearly constant frequency can be assigned to perpendicular standing spin waves with an increasing number of nodal planes across film thickness~\cite{Supp}. On the other hand, the lowest frequency branch clearly shows a negative group velocity (frequency decreases as wave-vector increases) for wave vector above a few rad/µm, which fits very well with the dispersion relation calculated for the stripe precursor mode (red line). This negative velocity can appear surprising at first glance since spin-waves in the Damon-Eshbach configuration normally have positive group velocity~\cite{14}. However, it was already observed in the presence of a perpendicular magnetic anisotropy~\cite{18} and it finds a natural explanation here: The perpendicular magnetic anisotropy favors the out-of plane component of the magnetization precession with respect to the in-plane one, which allows for a certain degree of dipolar field cancellation at sufficiently short length-scale. Decreasing the field to $14~$mT, \textit{i.e.} about $2~$mT above stripe nucleation, leads to a clear frequency decrease [Fig. 2(d)], which can be extrapolated to a perfect softening at ($k_c,H_c$).
Let us now examine the spin-wave dispersion below nucleation. Symbols in Fig. 3(a) show the positions of the Brillouin light scattering peaks measured at $7~$mT (see raw data in Fig. S5 in the Supplementary Information~\cite{Supp}).  We distinguish clearly  two branches. The frequency of the bottom one decreases rapidly as a function of wavenumber down to an extrapolate $f\sim 0$ at $k_c$. The frequency of the top one decreases much slower and extrapolates to a value of  about $3.5~$GHz at $k_c$. To help interpret these observations, we have performed mumax3 finite difference micromagnetic simulations~\cite{19} of spin-wave propagation. Fig. 3(a)-(c) show color plots of the amplitude spectral density obtained upon Fourier transforming the spatio-temporal evolution of the surface magnetization following a localized pulse excitation (see~\cite{Supp}) for field values of 7, 10 and $11.7~$mT, respectively. Right below nucleation [Fig. 3(c)], one distinguishes a secondary branch with a non-zero minimum frequency emerging from the characteristic $(k,f)=(k_c,0)$ cusp. Upon reducing further the field, the minimum frequency of this secondary branch gradually increases, while the main branch remains soft [Fig. 3(b)]. These two branches account very well for the measured inelastic peaks positions [Fig. 3(a)]. This phenomenology can be understood as the mesoscopic counterpart of the one occurring at the microscopic level for charge density waves~\cite{20} and incommensurate displacive phases~\cite{7} whose nucleation is also described by the softening of a dynamic mode which splits into amplitude and phase modes upon symmetry breaking. Fig. 3(d,e) show maps of the out-of-plane component of the dynamic magnetization $m_y(x,y)$ at $k_c$ for these two branches, together with the distribution of the transverse static magnetization, sketched as a vector plot. One clearly recognizes two similar patterns phase-shifted by $\pi/2$. For the zero frequency mode, the antinodes of the dynamic magnetization are aligned with the nodes of the static distribution [Fig. 3(d)]. In contrast, for the non-zero frequency mode, dynamic and static antinodes are aligned with each other [Fig. 3(e)]. The evolution with respect to the spectrum above saturation is explained as follows: The phase transition being second order, the overall spin wave spectrum changes smoothly upon stripe nucleation. The soft spin waves actually adapt in the form of intensity modulations in phase or in quadrature with respect to the nucleated texture.  

%Fig3
\begin{figure}
	\centering
	\includegraphics[width=1\linewidth]{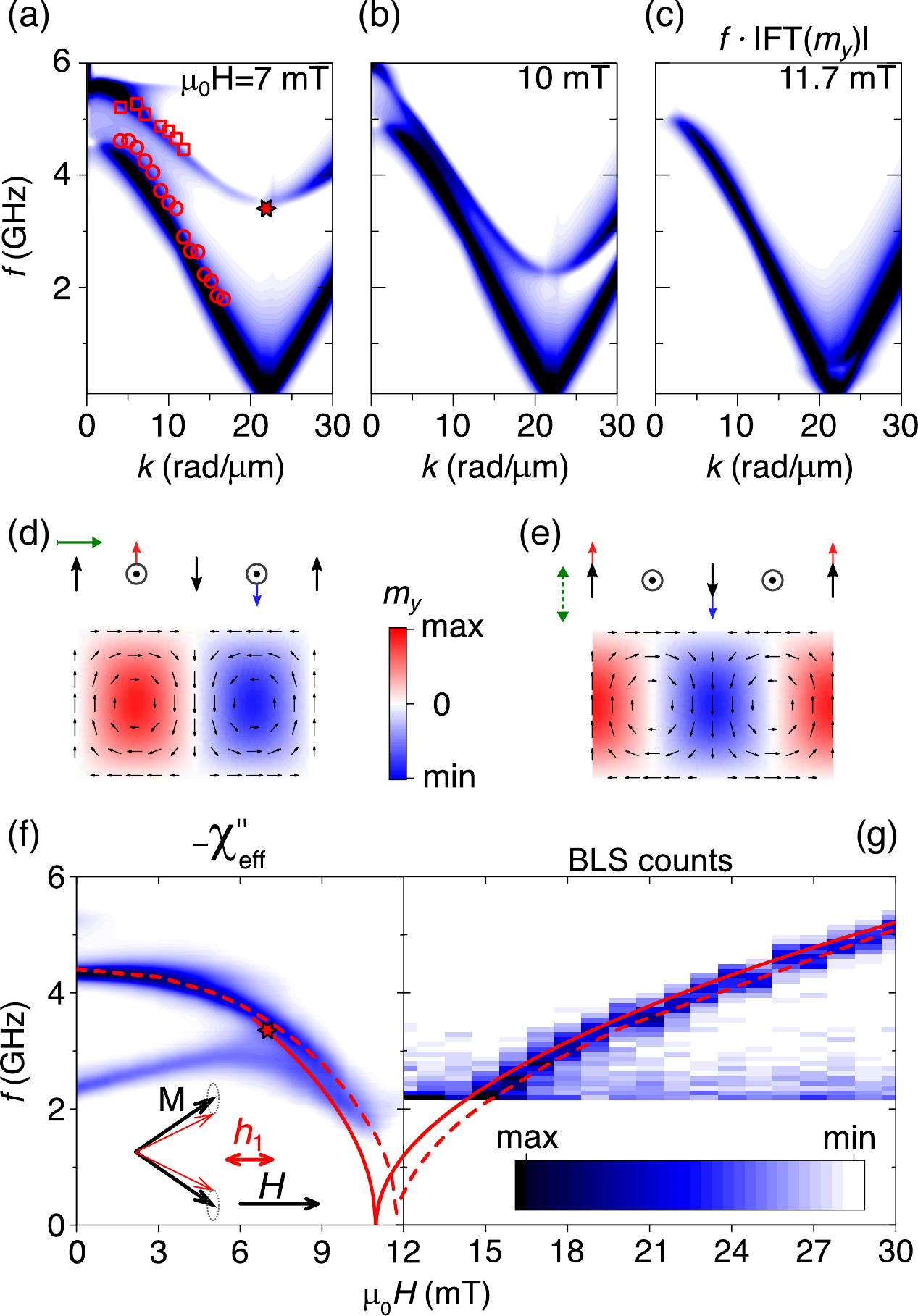}
	\caption{(a,b,c) Color plot of the simulated spin-wave spectrum amplitude density as function of wave-vector and frequency, for a field of 7, 10 and $11.7~$mT, respectively. In panel (a), the positions of the measured Brillouin light scattering peaks are reported as circles and squares and the Higgs mode frequency determined analytically is shown as a star. (d) Color plot of the distribution of the out-of-plane component of the dynamic magnetization for the Goldstone mode [$k= k_c$, $f=0.1~$GHz in panel (b)]. The vector plot shows the distribution of the transverse magnetization of the underlying static stripe modulation. (e) Same for the Higgs mode [$k= k_c$, $f=3.5~$GHz in panel (b)]. The green arrows sketch the motion of the stripe structure, namely a rigid displacement and an amplitude oscillation in panels (d) and (e), respectively. (f) Color plot of the microwave absorption measured in the stripe phase as function of field and frequency, for the longitudinal pumping geometry sketched in the inset. (g) Color plot of the Brillouin light scattering intensity measured in the saturated phase as function of field and frequency for a transferred wave-vector $k_c=21~$rad/$\mu$m. The solid and dashed lines in panels (f) and (g) show the soft/Higgs mode frequency extracted from our analytical approach (see details in the text and the Supplementary Information~\cite{Supp}) and from micromagnetic simulations, respectively.  The (min,max) values for color plots are (4,20), (-1,-18) $\times 10^{-3}$ and (100,300) for panels (a-c), (f) and (g), respectively.}\label{Fig_3}
\end{figure}

We shall now identify these two types of modulations with the Goldstone and Higgs modes of the stripe texture.  According to Fig. 1(a), in the low symmetry phase, one should distinguish phase oscillations $\delta=\delta_0 + \delta_1 \text{cos}(\omega t)$ and amplitude oscillations $|\psi| =\psi_0 +\psi_1 \text{cos}(\omega t)$. As these time-oscillations occur around an equilibrium which is oscillating in space (e.g. $M_y(x,0)\propto \text{cos}\left[k_c(x-\delta_0)\right]$), they correspond to non-zero wavenumber spin waves. More precisely, the dynamic magnetization profiles  (e.g. $\delta_1 \frac{\partial M_y}{\partial x} (x,0) \propto \text{sin}\left[k_c(x-\delta_0)\right]$ and $\psi_1 M_y (x,0) \propto \text{cos}\left[k_c(x-\delta_0)\right]$) can be interpreted as two standing wave patterns formed by the interference between counterpropagating spin waves with $k = \pm k_c$ and a well-defined phase difference of 0 or $\pi/2$ with respect to the equilibrium modulation. This corresponds exactly to the modal distributions of Fig. 3(d,e), to be identified with the Goldstone and Higgs modes, respectively. The zero frequency of the former is associated with the translation invariance of the whole stripe texture: the total energy is exactly the same whatever the value of the lateral shift $\delta$ in Fig. 1(d), as already noticed by~\cite{18}. The non-zero frequency of the Higgs mode arises from the finite curvature of the energy potential along the radial direction. We can evaluate this frequency via a suitable extension of the description of the dynamics above saturation~\cite{7,10}.  From the expression of the Landau potential, it can be shown that the positive curvature around the stable equilibrium value $\psi_0=\sqrt{\frac{-a}{2\,b}}$ is related to the negative curvature around the unstable equilibrium value $\psi=0$, namely $\left(\frac{\partial^2 E}{\partial |\psi|^2 } \right)_{\psi_0}=-2 \left(\frac{\partial^2 E}{\partial |\psi|^2 } \right)_{0}$. Then, we can write the frequency of the amplitude mode as $\omega \propto \sqrt{\frac{\partial^2 E}{\partial |\psi|^2} \frac{\partial^2 E}{\partial \tau^2} }$ where $\frac{\partial^2 E}{\partial \tau^2}$   is a ``kinetic'' term accounting for the extra energy generated by magnetization precession (Fig. S3 in~\cite{Supp}). This is essentially a strong demagnetizing contribution [Fig. 2(a)] which does not depend on the subtle energy balance that governs nucleation. It can therefore be assumed to be the same at $|\psi| = 0$ and $\psi_0$. Accordingly, we obtain $\omega|_{\psi_0}=-i\sqrt{2}\omega|_0$, which allows us to relate the frequency of the Higgs mode in the low symmetry phase to the growth rate of the unstable mode in a fictitious high symmetry state below nucleation. Using the growth rate calculated from our spin wave ansatz [bottom panel in Fig. 2(b)], we obtain the value shown as a star in Fig. 3(a), in good agreement with the numerical simulations.

To characterize directly the approach towards the critical point from both sides, we finally combine two techniques [Fig. 3(f,g)]. The mode softening in the saturated phase ($H>H_c$) is followed by Brillouin light scattering, placing ourselves exactly at $k_c$ [Fig. 3(g)]. There, one distinguishes clearly a gradual drop which follows precisely the characteristic softening predicted for the precursor mode by our analytical approach (solid line) or by micromagnetic simulations (dashed line). This technique becomes less efficient in the stripe phase because of the dephasing induced by inhomogenities of the stripe phase across the several tens of $\mu$m of the focal spot of the laser. This results in a sizeable drop of the light scattering signal at high wavenumber (Fig. S5(b) in~\cite{Supp}). Rather, we resort to another technique able to probe the stripe texture in a scalar way~\cite{21} (\textit{i.e.} regardless the phase of the nucleated texture~\cite{6}) namely ferromagnetic resonance under longitudinal pumping. The measurement configuration is shown in the inset of Fig. 3(f). The film is placed on top of a broadband transmission line (see Fig. S6(a) in~\cite{Supp}), which generates a (mostly in-plane and homogeneous) microwave magnetic field $\bm{h}_1$, the static field $\bm{H}$ being oriented parallel to it. This can be viewed as an analog of the lattice depth modulation technique used in cold atoms systems~\cite{9}: During a microwave cycle, the pumping field alternatively increases and decreases the total external field, which translates into an oscillation of the Zeeman energy and, in turn, into an oscillation of the amplitude of the order parameter. Fig. 3(f) shows the imaginary part of the effective magnetic susceptibility $\chi_{\text{eff}}$ of the loaded transmission line (which is proportional to the microwave absorption coefficient) as function of both the microwave frequency and  the static field intensity. One recognizes clearly a strong absorption feature below $H_c$, with a frequency increasing from about $1.5~$GHz in good agreement with the frequency upturn predicted by our analytical approach (solid line) and simulations (dashed line)\cite{note_FMR}. This absorption is associated to the excitation torque $\bm{M}(x,y)\times \bm{h}_1$ which is zero in the saturated state but increases gradually below nucleation due to the transverse static components of the stripe modulation ($\bm{M}\perp \bm{h}_1$). This allows us to shed a new light onto previous experiments of ferromagnetic resonance in stripe domains, traditionally interpreted in terms of distinct domain- and domain-wall resonances~\cite{22,23,24,25}. Our analysis indicates that the mode probed by longitudinal pumping corresponds to a spin-wave already present in the saturated state at $k=k_c$ and made accessible to a $k = 0$ experiment by a Bragg scattering process induced by the nucleated texture.

\section{\label{sec:level1}Conclusion}
To conclude, we show that both the close-to-nucleation statics and the low-frequency dynamics of magnetic stripe domains are entirely determined by the specific behaviour of flux-closure Damon-Eshbach spin-waves around a certain wave vector $k_c$.  The evolution of the whole spin-wave dispersion upon the high to low-symmetry transition can be analyzed in universal terms invoking the softening of the low frequency spin-wave branch, its freezing in the form of the translation-symmetry-breaking static stripe modulation, and its subsequent splitting into a Goldstone and a Higgs branch. The identified intimate relationship between the statics and the dynamics of a magnetic texture is a generic feature that could be taken advantage of in future developments of magnonics~\cite{26,27,28,29,30}. More importantly, the described system constitutes a particularly simple and explicit implementation of the dynamics around symmetry-breaking phase transitions, paving the way for further exploration, including time-resolved imaging studies, extension to the non-linear regime, or the quest for a possible Higgs-Anderson mechanism for magnons.

\begin{acknowledgments}
We thank S. Cherifi-Hertel and H. Majjad for complementary magnetic imaging, T. Böttcher and M. Ruhwedel for support with the BLS measurements, and J.V. Kim, J. P. Adam, T. Devolder, J. Solano and G. DeLoubens for fruitful discussions. 

\vspace{0.5cm}

This work was funded primarily by the French National Research Agency (ANR) through the project SWANGATE (ANR-16-CE24-0027). We acknowledge the technical support of the STnano clean-room platform partly funded by the Interdisciplinary Thematic Institute QMat, as part of the ITI 2021-2028 program of the University of Strasbourg, CNRS and Inserm, IdEx Unistra (ANR 10 IDEX 0002), SFRI STRAT’US project (ANR 20 SFRI 0012) and ANR-17-EURE-0024 under the framework of the French Investments for the Future Program. We acknowledge support from the High Performance Computing Center of the University of Strasbourg which provides access to computing resources, partly funded by the Equipex Equip@Meso project (Programme Investissements d'Avenir) and the CPER Alsacalcul/Big Data. We also acknowledge financial support from Region Grand Est through its FRCR call (NanoTeraHertz and RaNGE projects), from the impact project LUE-N4S part of the French PIA project “Lorraine Université d’Excellence”, reference ANR-15IDEX-04-LUE and from the “FEDER-FSE Lorraine et Massif Vosges 2014–2020”, a European Union Program. MG and PP acknowledge funding by the Deutsche Forschungsgemeinschaft (DFG, German Research Foundation) - TRR 173 - 268565370 (“Spin + X”, Project B01) and Priority Programme SPP2137 Skyrmionics. 

\vspace{0.5cm}

M.Gr., D. S., Y.H. and M.B. designed the experiments. M. H. fabricated the magnetic films. K. A. and D. L. performed magnetometry and magnetic imaging. M. Ge. performed Brillouin light scattering measurements. M. Gr. performed ferromagnetic measurements and analysed the experimental data, with the help of M. Ge., P. P. and M. B. M. Gr. and M. B. developed the analytical model. D. S. performed the micromagnetic simulations with the help of Y. H. and M. Gr. M. Gr. and M. B. wrote the manuscript with the help of all authors.

\end{acknowledgments}



\begin{thebibliography}{0}
\bibitem{21} 
In this context scalar means that the probe is not sensitive to the phase of the texture. This is in contrast with a vectorial probe which measures a projection along a certain phase realization~\cite{6}, as Brillouin light scattering which involves a well-defined phase over the whole laser beam.
\end{thebibliography}
\end{document}